\documentclass[twocolumn,english]{revtex4}
\usepackage[T1]{fontenc}
\usepackage[latin9]{inputenc}
\usepackage{amssymb}
\usepackage{graphicx}

\makeatletter
\@ifundefined{textcolor}{}
{%
 \definecolor{BLACK}{gray}{0}
 \definecolor{WHITE}{gray}{1}
 \definecolor{RED}{rgb}{1,0,0}
 \definecolor{GREEN}{rgb}{0,1,0}
 \definecolor{BLUE}{rgb}{0,0,1}
 \definecolor{CYAN}{cmyk}{1,0,0,0}
 \definecolor{MAGENTA}{cmyk}{0,1,0,0}
 \definecolor{YELLOW}{cmyk}{0,0,1,0}
}

\makeatother

\usepackage{babel}
\begin{document}

\title{Smooth phase transition of energy equilibration in a springy Sinai
billiard}

\author{Kushal Shah}
\email{kushals@iiserb.ac.in}

\affiliation{Department of Electrical Engineering and Computer Science, Indian
Institute of Science Education and Research (IISER), Bhopal - 462066,
Madhya Pradesh, India.}
\begin{abstract}
Statistical equilibration of energies in a slow-fast system is a fundamental
open problem in physics. In a recent paper, it was shown that the
equilibration rate in a springy billiard can remain strictly positive
in the limit of vanishing mass ratio (of the particle and billiard
wall) when the frozen billiard has more than one ergodic components
{[}Proc. Natl. Acad. Sci. USA 114, E10514 (2017){]}. In this paper,
using the model of a springy Sinai billiard, it is shown that this
can happen even in the case where the frozen billiard has a single
ergodic component, but when the time of ergodization in the frozen
system is much longer than the time of equilibration. It is also shown
that as the size of the disc in the Sinai billiard is increased from
zero, thereby leading to a decrease in the time required for ergodization
in the frozen system, the system undergoes a smooth phase transition
in the equilibration rate dependence on mass ratio. 
\end{abstract}


\maketitle
Since the seminal work done by James C. Maxwell, Ludwig Boltzmann,
J. William Gibbs and few others in the 1800s, statistical physics
has made immense progress for equilibrium systems and many of the
laws proposed by statistical physicists have found applications in
diverse fields like information theory \cite{Merhav}, economics \cite{Sinha}
and now even in machine learning \cite{Hertz}. The most fundamental
connection between statistical mechanics and information theory is
that of entropy. Boltzmann and Gibbs gave the first probabilistic
description of entropy in the late 1800s, but it was Claude E. Shannon
who came up with a mathematically rigorous derivation for the entropy
formula in 1948 while studying communication systems \cite{Merhav}.
Since then, several other statistical mechanics concepts like the
maximum entropy principle, random matrix theory and even spin glasses
have found many applications in information theory and communication
systems \cite{Merhav}. Applications of statistical physics and nonlinear
dynamics concepts in economics have given birth to a whole field of
econophysics, which attempts at coming up with various mathematical
models to explain economic phenomenon and predict its future course
\cite{Sinha}. Machine learning is not a new concept, but recent advances
in processing power of computer chips has made it possible for humans
to harness its power on reasonable time scales, thereby leading to
a complete paradigm shift in the way we think of computations and
algorithms. And perhaps not surprisingly, several statistical physics
concepts like the Boltzmann-Gibbs distribution and stochastic dynamics
have found to be very useful in building and explaining models of
machine learning \cite{Hertz}.

Although lot of progress has been made in the field of non-equilibrium
statistical mechanics, most of the well established laws of statistical
mechanics mainly pertain to systems under equilibrium. One of the
fundamental open problems in statistical physics is about how dynamical
systems actually reach this state of statistical equilibrium! A partial
answer is provided by the ergodic hypothesis which states that all
accessible microstates of a given system are equiprobable over sufficiently
long periods of time \cite{Dorfman,Moore}. However, there are very
few dynamical systems which have been actually proven to be ergodic
\cite{Simanyi,Sinai1970,Sinai1987,RomKedar,Rapoport,Kaplan}. And
even for ergodic systems, the time required for ergodization may be
so long that it may be practically irrelevant. In the case of slow-fast
ergodic systems, it has also been shown that there are adiabatic invariants
which can prevent equilibration of the full system over very long
periods of time \cite{Neishtadt,Wright}. Hence, there is a theoretical
as well as practical need to understand the equilibration properties
of systems from the dynamical perspective.

\begin{figure}
\begin{centering}
\includegraphics[scale=0.35]{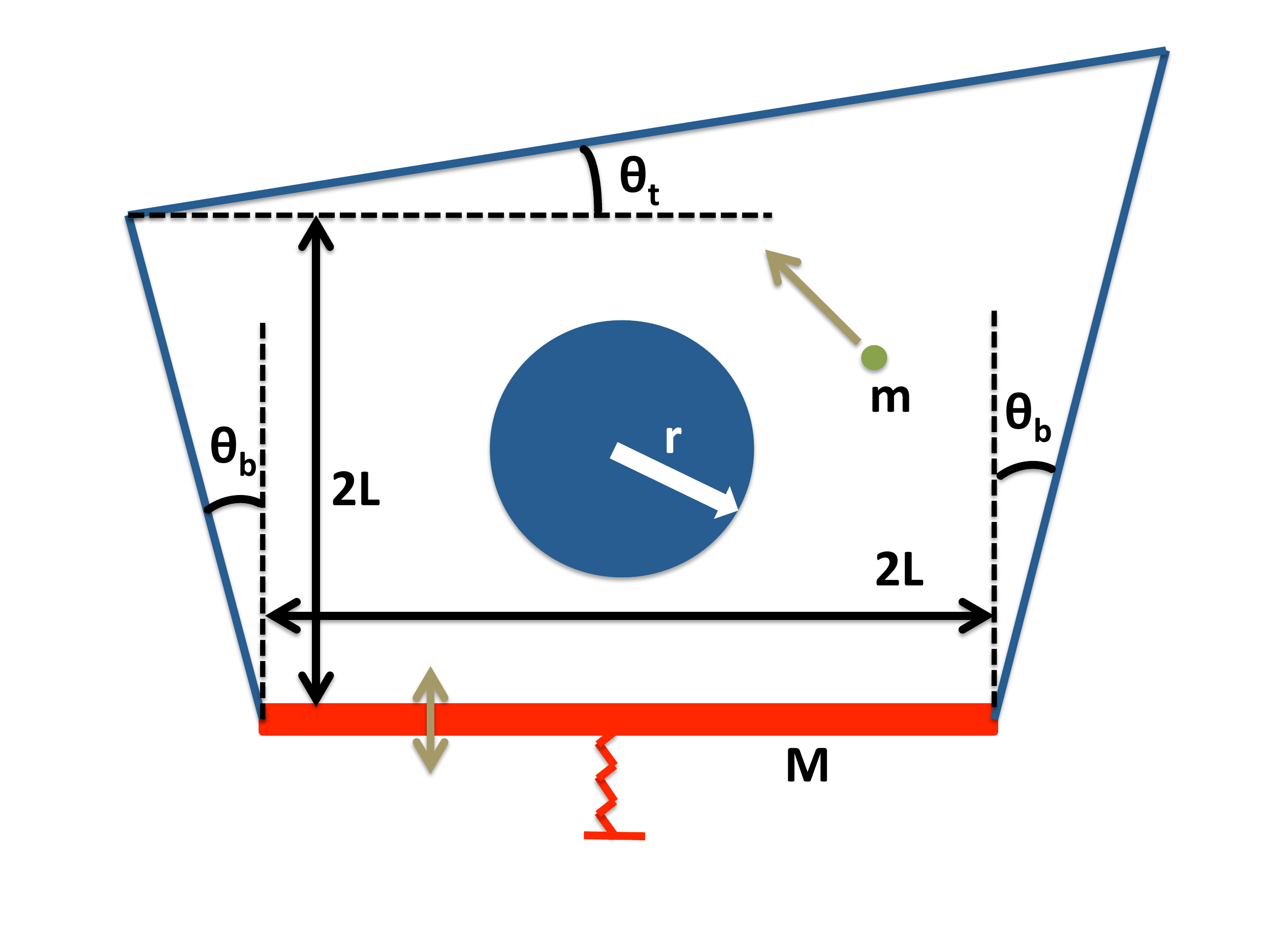}
\par\end{centering}
\caption{Springy Sinai billiard, which consists of a particle of mass, $m\ll1$,
moving within the billiard boundaries undergoing elastic reflections
at each collision with the boundaries (including the disc in between).
The bottom wall of the billiard has a mass, $M=1$, and is attached
to a spring such that its natural frequency of oscillations is $\omega=1$.
When the disc radius is zero, this becomes a polygon (which is non-ergodic
if all its angles are rational multiples of $\pi$) and for a non-zero
disc radius, this becomes a Sinai billiard which is known to be hyperbolic.
It is this transition in the billiard properties as we change the
disc radius that leads to the phase transition that is demonstrated
in this paper. The parameter values chosen are $L=2$, $\theta_{b}=\pi\big/18=\theta_{t}$,
$0\le r\le1$, $M=1$ and $4\times10^{-6}\le m\le12\times10^{-6}$.
\label{fig:Slanted-Sinai-billiard}}
\end{figure}

In a recent paper, it has been shown that equilibration of energies
can be achieved in slow-fast systems on reasonable time scales if
the frozen system has more than one ergodic components \cite{Shah2017}.
This was numerically demonstrated by studying the dynamics of a point-like
particle of small finite mass, $m$, in a springy billiard where one
of the walls is massive, $M\gg m$, and is connected to a linear spring.
Three different springy billiards were studied in that paper : springy
barred rectangle, springy mushroom and springy stadium. The total
energy of the springy billiard system is conserved in each case since
its an autonomous system. It was found that the partial energies of
the particle, $E_{p}$, and the massive billiard wall, $E_{b}$, reached
a state of equipartition and equilibration asymptotically with time
for all the three springy billiards when the mass ratio $m\big/M$
was non-zero. However, in the limit of vanishing mass ratio, only
the springy barred rectangle and springy mushroom retained a non-zero
equilibration rate, whereas the equilibration rate for the springy
stadium went to zero. It was shown that this difference in behavior
can be explained through a mathematical model by taking into account
the fact that the springy barred rectangle and springy mushroom have
more than one ergodic components in the frozen state (called VFS systems,
variable partition of the fast subspace), and the springy stadium
has only one ergodic component (called EFS systems, ergodic fast subsystem
for almost all values of the slow variables). However, one similarity
between the springy mushroom and the springy stadium was that in both
the cases, the equilibration rate varied with mass as $\sqrt{m\big/M}$,
which is same as what was predicted earlier for the case of uniformly
hyperbolic systems \cite{Wright,Neishtadt}. For the case of the springy
barred rectangle, the equilibration rate was found to be independent
of the mass ratio as also predicted by the mathematical model \cite{Shah2017}.
Though the above result is expected to hold for systems in general
which can clearly be classified as being VFS or EFS, the behavior
of systems which are in between can be lot more interesting. One example
of such a system is the springy Sinai billiard as shown in Fig. \ref{fig:Slanted-Sinai-billiard},
which essentially consists of a circular disc within a trapezium. 

In this paper, the springy Sinai billiard has been studied and found
to have several very interesting properties so far not reported in
any other slow-fast system. Most importantly, it has been found that
this billiard can have a non-zero value of the equilibration rate
in the limit of vanishing mass ratio even when the frozen system has
a single ergodic component. When the disc radius is increased, the
equilibration rate tends to zero in the limit of vanishing mass ratio
as is typical for EFS systems. Hence, this billiard is found to undergo
a smooth phase transition from a VFS-like system to an EFS-like system
as we increase the disc radius. Interestingly, for low values of the
disc radius, the equilibration rate dependence on the mass ratio is
also found to be non-monotonic. This hints at the possibility that
this dynamical system can be a very good candidate for the discovery
of interesting dynamical properties not commonly found in other springy
billiards. 

\begin{figure}
\begin{centering}
\includegraphics[scale=0.35]{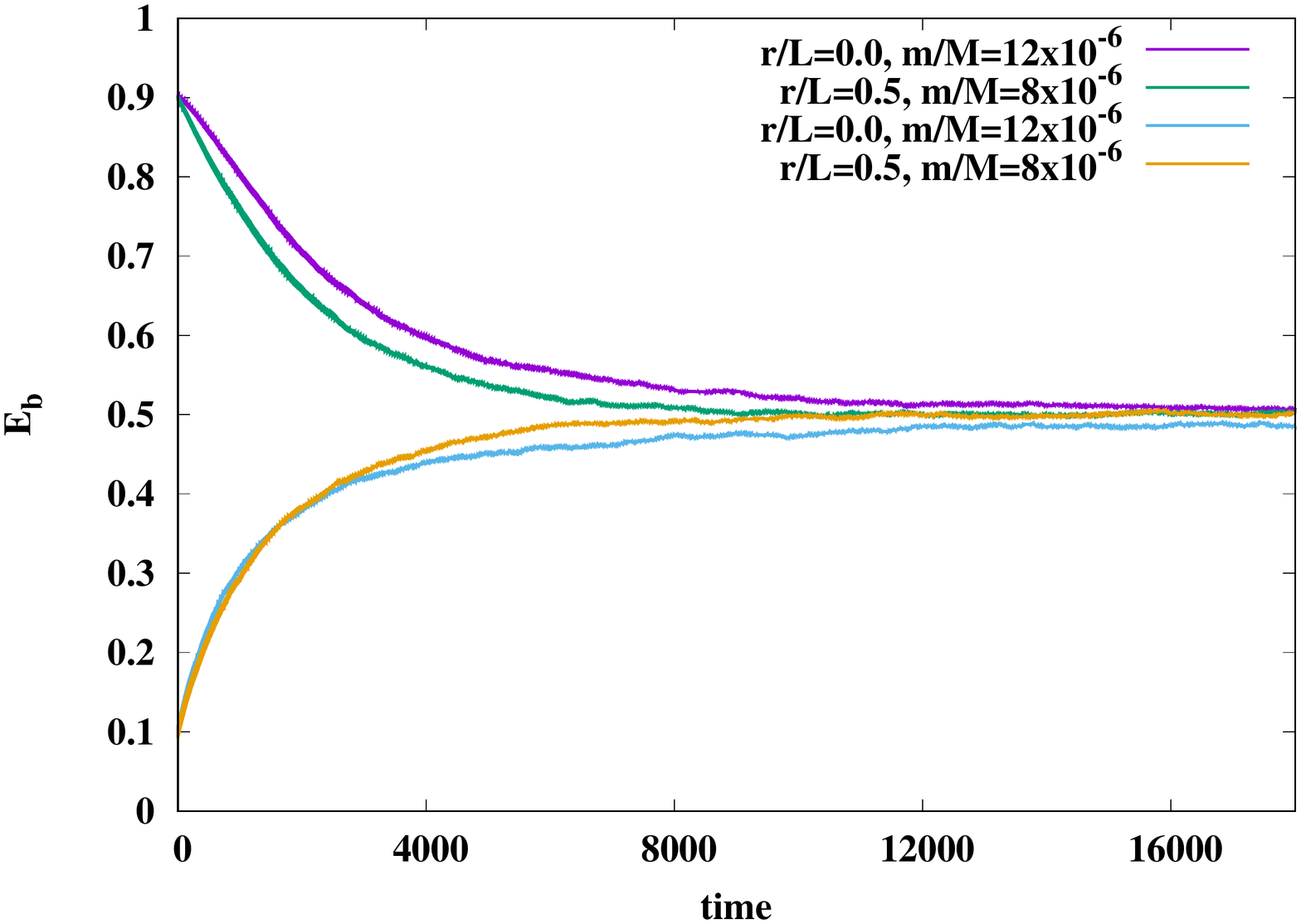}
\par\end{centering}
\caption{Variation of energy of the oscillating bar, $E_{b}$, with time for
few values of $r\big/L$ and $m\big/M$. As can be seen in this figure,
the bar energy tends towards the equilibration at $E_{b}=0.5$ irrespective
of its starting value. \label{fig:Variation-of-energy}}
\end{figure}

\begin{figure}
\begin{centering}
\includegraphics[scale=0.35]{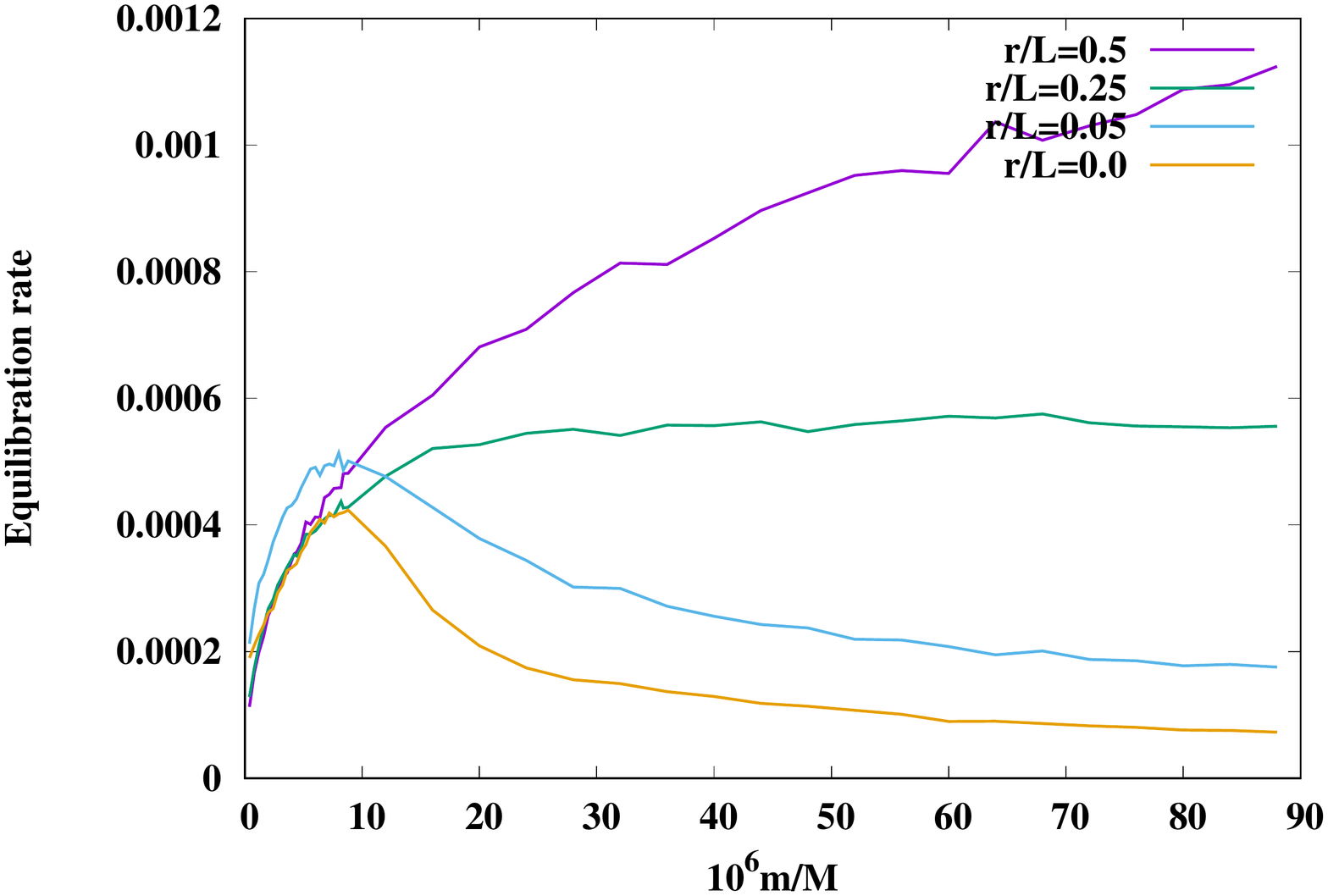}
\par\end{centering}
\caption{Equilibration rate, $\gamma$, dependence on the mass ratio, $m\big/M$
for few values of the disc radius, $r\big/L$ for $E_{b}\left(0\right)=0.9$.
For low values of $r\big/L$, $\gamma$ is non-monotonic in $m\big/M$
and has a non-zero value in the limit of vanishing mass ratio. However,
as the value of $r\big/L$ crosses a certain critical value, $\gamma$
shows is monotonic in $m\big/M$ and tends towards a zero limiting
value for large enough radius. For intermediate values of $r\big/L\sim0.25$,
the value of $\gamma$ becomes independent of the mass ratio beyond
a certain threshold, which is similar to the behavior observed in
the springy barred rectangle \cite{Shah2017}. \label{fig:game-versus-m-Eb9}}
\end{figure}

\begin{figure}
\begin{centering}
\includegraphics[scale=0.3]{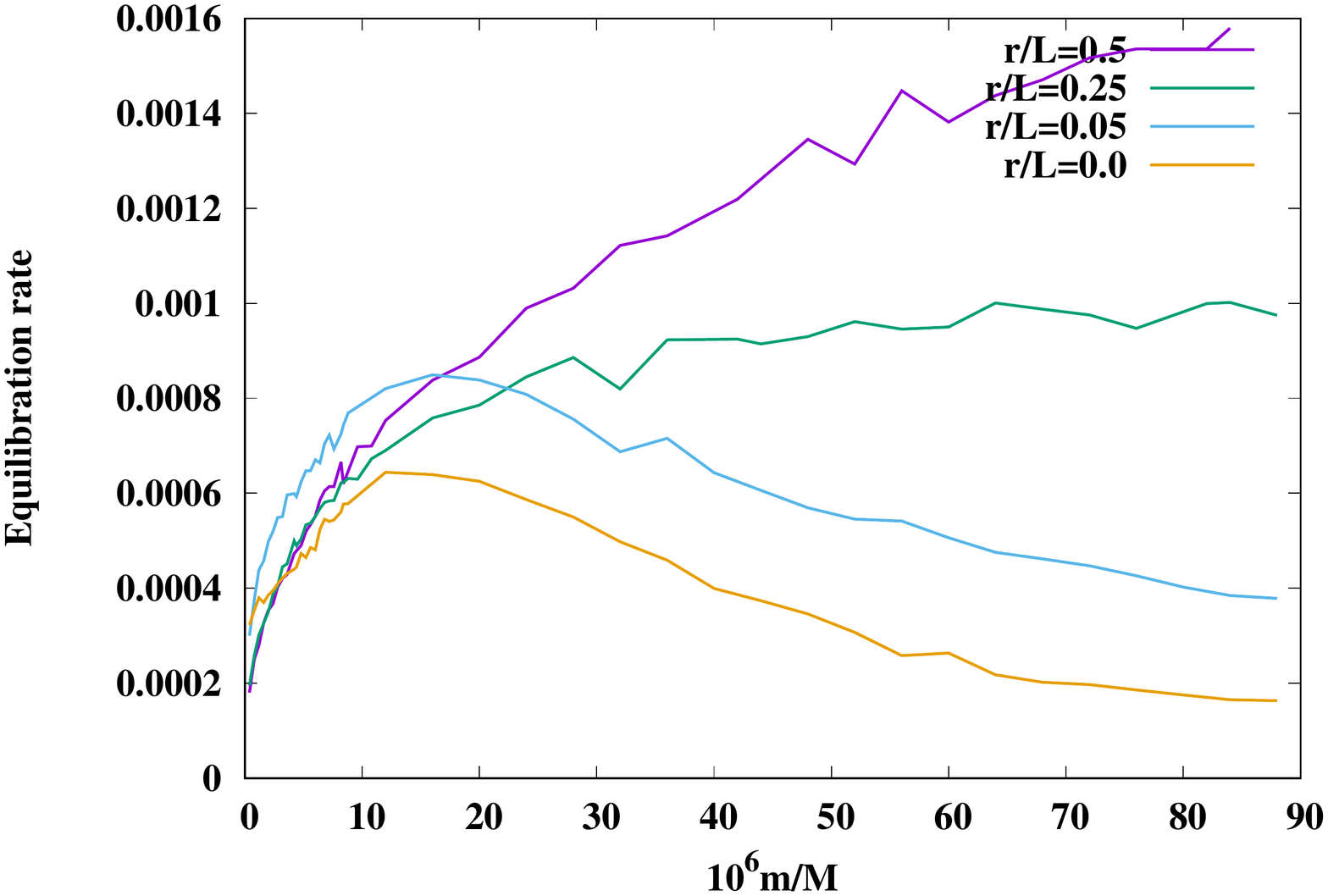}
\par\end{centering}
\caption{Equilibration rate, $\gamma$, dependence on the mass ratio, $m\big/M$
for few values of the disc radius, $r\big/L$ for $E_{b}\left(0\right)=0.1$.
The behavior is found to be qualitatively similar to that for $E_{b}\left(0\right)=0.9$
as shown in Fig. \ref{fig:game-versus-m-Eb9}. For low values of $r\big/L$,
$\gamma$ is non-monotonic in $m\big/M$ and has a non-zero value
in the limit of vanishing mass ratio. However, as the value of $r\big/L$
crosses a certain critical value, $\gamma$ shows is monotonic in
$m\big/M$ and tends towards a zero limiting value for large enough
radius.\label{fig:game-versus-m-Eb1}}
\end{figure}

The conventional Sinai billiard consists of a circular disc at the
center of a rectangle and in fact, was one of the first dynamical
billiards to be shown to be hyperbolic \cite{Sinai1970,Sinai1987,Bunimovich}.
If one of the walls of this billiard is attached to a linear spring,
it is expected to show similar equilibration properties as that of
the springy Stadium \cite{Shah2017}. In the limit of vanishing disc
radius, the springy Sinai billiard is reduced to a springy rectangular
billiard which is known to be integrable \cite{Gutkin}, and hence
does not have any equilibration of energies even for a non-zero mass
ratio \cite{Shah2017}. In an integrable billiard attached to a spring,
the partial energies of the particle and the oscillating bar keep
varying periodically about a certain average without reaching equilibration
\cite{Shah2017,Wright,Neishtadt}. 

As shown in Fig. \ref{fig:Slanted-Sinai-billiard}, the walls of the
springy Sinai billiard have been made slanted in this work to create
a trapezium, which is known to be non-integrable \cite{Richens,Shah2011,Gutkin}
and hence, is expected to have equilibration of energies asymptotically
even in the limit of zero mass ratio. In fact, among all possible
varieties of polygons, it is known that only four are integrable :
rectangle, equilateral triangle, right angled triangle with other
two angles $\pi\big/4$, and right angled triangle with other two
angles $\pi\big/3$ and $\pi\big/6$. If the angles of the non-integrable
polygon are rational multiples of $\pi$, then the billiard is also
known to be non-ergodic and hence, is expected to show a behavior
similar to that of VFS systems \cite{Shah2017}. Very little is known
about the ergodic properties of polygons with angle(s) which are irrational
multiple(s) of $\pi$ and it is one of the most important open questions
in the field of dynamical billiards \cite{Bobok,Gutkin1995,Gutkin2003,Dingle}.
Polygonal billiards are also of importance in the study of quantum
mechanics, and can have very interesting solutions for the quantum
energy levels with important implications in the field of quantum
chaos \cite{Richens}.

The rational trapezium with a circular disc in between is a very interesting
system. When the radius of the disc is non-zero, the system is known
to be ergodic, and the time-scale of ergodization decreases as the
disc radius increases \cite{Sinai1987}. Hence, when the disc radius
of this springy Sinai billiard is large enough, it is expected to
behave like an EFS system, but it is apriori not clear what may happen
when the disc radius is small enough. This is because for small disc
radius, the dynamical properties of the non-ergodic non-integrable
trapezium may become more dominant. Hence, it might behave like an
EFS system for all non-zero values of the disc radius, or might undergo
a phase transition to a VFS-like system when the disc radius is small
enough. This question is numerically studied in this paper and the
latter possibility is found to be true. The springy Sinai billiard
indeed undergoes a smooth phase transition from an EFS system to a
VFS-like system as the disc radius is decreased.

The springy Sinai billiard shown in Fig. \ref{fig:Slanted-Sinai-billiard}
consists of a circular disc of radius, $r$, contained within a trapezium.
In this paper, the values taken are $L=2$, $\theta_{b}=\pi\big/18=\theta_{t}$,
$0\le r\le1$, $M=1$ and $4\times10^{-7}\lesssim m\lesssim9\times10^{-5}$.
Simulations were performed for other values of $\theta_{t},\theta_{b}$
and qualitatively similar results were found as those reported in
this paper. The spring attached to the massive billiard boundary has
a spring constant such that its angular frequency of oscillations
is $\omega=2\pi$. Numerical simulations are performed for an ensemble
of 10000 particles using the same algorithm described in \cite{Shah2017}.
The particle moves inside the billiard in straight lines and undergoes
elastic collisions at the boundaries. The particle velocity after
collision with a static wall is simply given by the law of elastic
collisions, where the angle of incidence is equal to the angle of
reflection. In this case, the particle velocity only undergoes a change
in direction and its speed remains the same. When the particle undergoes
elastic collisions with the oscillating bar, the time and position
of collision is calculated using a combination of bisection and Newton
method. The equilibration rate is estimated using a linear least squares
fit of $log\left|E_{b}-0.5\right|$ over a time-interval in which
the bar energy, $E_{b}$, changes by a factor of $e$. In each simulation,
there is only a single particle in the springy billiard and then an
average is taken over 10000 different randomly chosen initial conditions.
The total energy of the system stays constant at $E=1$, which is
the sum of the particle kinetic energy, $E_{p}$, and the energy of
the oscillating bar, $E_{b}$. The energy $E_{b}$ is a sum of the
kinetic energy of the bar and the potential energy of the attached
spring. There is an exchange of energy between $E_{p}$ and $E_{b}$
each time there is a collision between the oscillating bar and the
particle. In between such collisions, the values of $E_{p}$ and $E_{b}$
remain unchanged.

Figure \ref{fig:Variation-of-energy} shows the plot of the bar energy,
$E_{b}$, with time for few values of $r\big/L$ and $m\big/M$ each.
As can be clearly seen, the bar energy reaches its equilibrium value
of $0.5$ in all cases, which is the expected behavior for billiards
which are non-integrable \cite{Shah2017}. The equilibration proceeds
approximately as an exponential function, and so $E_{b}$ can be written
as 
\begin{equation}
E_{b}\left(t\right)\approx0.5+\left[E_{b}\left(0\right)-0.5\right]e^{-\gamma t}\label{eq:Eb-exp}
\end{equation}
where $\gamma$ is the equilibration rate and depends on $E_{b}\left(0\right)$,
$m\big/M$ as well as the billiard parameters. In this paper, we have
kept all other billiard parameters fixed, except for the disc radius,
$r$.

Figure \ref{fig:game-versus-m-Eb9} shows a plot of the equilibration
rate, $\gamma$, versus the mass ratio, $m\big/M$, for a few values
of $r\big/L$ when $E_{b}\left(0\right)=0.9$. As can be seen, for
larger values of $r\big/L$, the equilibration rate is increases with
an increase in $m\big/M$ as a power-law and tends to zero in the
limit of vanishing mass ratio as is expected of EFS systems. However,
for lower values of $r\big/L$, $\gamma$ becomes non-monotonic and
has a non-zero value in the limit of vanishing mass ratio, which is
typical of VFS systems. For some values of the disc radius around
$r\big/L\sim0.12$, we also see that the equilibration rate is independent
of the mass ratio, which is similar to the behavior found for the
springy barred rectangle \cite{Shah2017}. A qualitatively similar
behavior is observed when $E_{b}\left(0\right)=0.1$ as shown in Fig.
\ref{fig:game-versus-m-Eb1}. Hence, it is reasonable to conclude
that this is typical behavior of the springy Sinai billiard. This
result is significant since it is usually believed that ergodic systems
should not have equilibration of energies in the limit of vanishing
mass ratio \cite{Wright,Neishtadt}. But this result shows that if
the ergodicity is weak, then even ergodic systems can have equilibration
of energies in this limit.

These results are particularly relevant to practical systems, since
most of them are neither strictly VFS or EFS and actually fall somewhere
in between, in the same sense as most real systems have a mixed phase
space. Hence, most real systems are expected to show this kind of
a smooth phase transition as the relevant parameters are varied. The
criteria for observing a similar behavior is that for some set of
parameters, the system should become strongly ergodic and for some
other set of parameters, the system should become strongly non-ergodic,
while remaining non-integrable for all parameter values. The simultaneous
requirement of non-ergodicity and non-integrability for some parameter
values is important since in the springy Sinai billiard shown in Fig.
\ref{fig:Slanted-Sinai-billiard}, if the bounding polygon is a rectangle
(non-ergodic, but integrable) instead of a trapezium, then the phase
transition will not be observed. This is because there is no equilibration
of energies in an integrable system with a linear spring and the bar
energy keeps oscillating about a certain mean value for all time.
However, there might be interesting equilibration effects even in
integrable billiards when the spring becomes nonlinear \cite{Schmelcher}. 

These results can also have very interesting implications in the study
of Fermi acceleration in dynamical billiards \cite{Shah2010,Shah2011,Gelfreich2008,GelfreichPRL}.
Fermi acceleration is the study of particle dynamics within billiards
in a similar manner as studied in this work, with the only difference
that there the billiard wall is infinitely massive and, hence, the
total energy of the system is not a conserved quantity. So, instead
of equilibration, what is observed is an unbounded increase of energy
of the particle ensemble with time if the underlying billiard is non-integrable.
In most such systems studied so far, this energy growth has been found
to be either exponential or quadratic in time. The prevailing understanding
is that the energy growth rate is quadratic-in-time if the underlying
frozen billiard is ergodic \cite{Gelfreich2008} and exponential-in-time
if it is non-ergodic \cite{Shah2011,Shah2010}. One of the open questions
in this area is whether polygons, which are known to be pseudo-integrable
\cite{Richens}, in general have an exponential growth of energy or
not. Some indirect evidence has been found which indicates an exponential-in-time
growth of energy in polygons \cite{Shah2010}, but it is not yet well
established. And as shown in \cite{Shah2017}, there is a strong connection
between equilibration rates in springy billiards in the limit of vanishing
mass ratio and exponential acceleration when the same system is studied
in the context of Fermi acceleration. The results reported in this
paper provide more evidence to support the case of exponential energy
growth in polygons in general. A more careful study of these models
might also be helpful in figuring out whether irrational polygons
are ergodic or not! 

This work primarily has three limitations, which can serve as fruitful
directions for future work. Firstly, no mathematical model could be
proposed in this paper to predict the equilibration rate in the limit
of vanishing mass ratio as was done for the springy barred rectangle
and the springy mushroom in \cite{Shah2017}. Proposing such a mathematical
model requires clear knowledge of the ergodic partitions of the frozen
system and the probability of jumping across these partitions. Both
these pieces of information were available for the springy billiards
studied in \cite{Shah2017}, but as of now, are not available for
the springy polygonal billiard mainly because, in this case, the ergodic
components are not that well separated as compared to the springy
barred rectangle or the springy mushroom. In the case of the springy
polygon, the ergodic components are intricately intertwined with each
other, and hence predicting the limiting equilibration rate will perhaps
require a completely new approach compared to that adopted in \cite{Shah2017}.
Secondly, it is not clear why the equilibration rate, $\gamma$, is
non-monotonic with respect to the mass ratio for low values of $r\big/L$.
Perhaps there is some kind of resonance phenomenon taking place for
certain values of $m\big/M$ for lower values of the disc radius,
which leads to a maximization of the equilibration rate. Thirdly,
the functional dependence of the equilibration rate on $m\big/M$
is unclear for lower values of the disc radius. This information is
required so as to be able to predict the value of the equilibration
rate in the limit of vanishing mass ratio. We can graphically see
that this limiting value is most likely non-zero, but a proper empirical
estimation is needed in order to be sure. 
\begin{acknowledgments}
This work was financially supported by a research grant from the Science
and Engineering Research Board {[}SERB{]}, Government of India (File
No. EMR/2016/001196).
\end{acknowledgments}

\end{document}